\documentclass[aps,prb,preprint,groupedaddress]{revtex4-2}
\usepackage{graphicx}
\usepackage{epstopdf}
\usepackage{color,soul}
\usepackage{hyperref}
\usepackage{lineno}
\usepackage{amsmath}
\usepackage{mathtools }
\usepackage[normalem]{ulem}
\newcommand{\comment}[1]{}
\begin{document}
\title{Temperature-dependent Fermi surface probed by Shubnikov-de Haas oscillations in topological semimetal candidates DyBi and HoBi}
\author{Paulina Nowakowska}
\author{Orest Pavlosiuk$^*$}
\author{Piotr Wiśniewski}
\author{Dariusz Kaczorowski}
\affiliation{Institute of Low Temperature and Structure Research, Polish Academy of Sciences, Ok\'{o}lna 2, 50-422 Wroc{\l}aw, Poland}
\affiliation{$^*$~Corresponding author: o.pavlosiuk@intibs.pl}
%
\begin{abstract}
Rare earth-based monopnictides are among the most intensively studied groups of materials in which extremely large  magnetoresistance has been observed. 
This study explores magnetotransport properties of two representatives of this group, DyBi and HoBi. 
The extreme magnetoresistance is discovered in DyBi and confirmed in HoBi.  
At $T=2$\,K and in $B=14$\,T for both compounds, magnetoresistance reaches the order of magnitude of $10^4\%$. 
For both materials, standard Kohler's rule is obeyed only in the temperature range from 50\,K to 300\,K.
At lower temperatures, extended Kohler's rule has to be invoked because carrier concentrations and mobilities strongly change with temperature and magnetic field. 
This is further proven by the observation of a quite rare temperature-dependence of oscillation frequencies in Shubnikov-de Haas effect.
Rate of this dependence clearly changes at N\'{e}el temperature, reminiscent of a novel magnetic band splitting.  
Multi-frequency character of the observed Shubnikov-de Haas oscillations points to the coexistence of electron- and hole-type Fermi pockets in both studied materials.
Overall, our results highlight correlation of temperature dependence of the Fermi surface with the magnetotransport properties of DyBi and HoBi.

\end{abstract}
\maketitle
\section{Introduction}

Rare-earth monopnictides ($REPn$, where $RE$ denotes one of the rare-earth elements and $Pn=$\,As, Sb or Bi) have attracted broad scientific and technological interest due to their exceptional physical properties and potential applications in solar cells,~\cite{Zide2006} thermoelectrics,~\cite{LIU201156} plasmonics,~\cite{Krivoy2018} and spintronics.~\cite{Schrunk2022}
In addition, the preparation of $RE$Sb thin films has been developed,~\cite{Inbar2022, Chatterjee2019a} which may be important from the point of view of future applications. 

Recently, interest in $REPn$ materials has surged due to the possible topologically non-trivial nature of their electronic bands.~\cite{Tafti2015}
In addition, unsaturated and extreme magnetoresistance (XMR), accompanied by magnetic field-induced low-temperature resistivity plateau, has been reported in many $REPn$ materials.~\cite{Pavlosiuk2017, Pavlosiuk2016, Niu2016, Vashist2019, Song2018b, Tafti2015}
At first, these phenomena in LaSb have been attributed to the existence of topologically non-trivial electronic states.~\cite{Tafti2015}
However, alternative explanation that has been postulated for many $REPn$ compounds is perfect or nearly-perfect compensation of charge carriers.~\cite{Ghimire2016,Kumar2016,Wu2019, Pavlosiuk2018,Pavlosiuk2017,Ye2018,Song2018b} 
In order to experimentally verify the presence of topologically non-trivial states in $REPn$ compounds, their electronic structure has been studied directly by angle-resolved photoemission spectroscopy (ARPES). 
In LaBi, for example, several groups have found the presence of topologically non-trivial states of different natures: nodal-lines,~\cite{Feng2018} surface states,~\cite{Jiang2018,Lou2016,Niu2016} and multiple Dirac cones.~\cite{Nayak2016}  
In addition, pressure-induced superconductivity has been reported for LaBi and YBi,~\cite{Xu2019, Tafti2016} which established them as possible topological superconductors, potentially hosting Majorana fermions.

The presence of a rare-earth element with partially filled $f$-shells in $REPn$ compounds provides even more fascinating physical properties. 
PrBi has been reported to exhibit nontrivial Berry phase and quadrupolar moment,~\cite{He2020} whereas topological quantum phase transition has been proposed in HoSb.~\cite{Zhang2021c}
Several antiferromagnets among $REPn$ family have been reported to demonstrate XMR.~\cite{Song2018b,Liang2018c,Wu2019,Tang2022}
Interestingly, it has been proposed that within $RE$Sb and $RE$Bi series topological phase transitions could occur close to SmSb and DyBi, respectively.~\cite{Duan2018}
Intriguing new magnetic splitting effect has recently been discovered in NdBi \cite{Schrunk2022} and other $RE$Bi compounds.~\cite{Kushnirenko2022} 
This splitting takes place below the N\'{e}el temperature ($T_N$) and leads to the appearance of new sheets of the Fermi surface, hole- and electron-like "magnetic" Fermi arcs, and it can lead to a change in the shape of the bulk Fermi pockets.
Interestingly, these arcs can be topologically non-trivial in nature.~\cite{Schrunk2022} 

We hypothesized that this behavior could be reflected in the properties of quantum oscillations (the appearance of new frequencies and the change in frequency with changing temperature), since the oscillation frequency is directly proportional to the cross-sectional area of the Fermi pocket.~\cite{Shoenberg1984}  
In NdBi, $T_N=24$\,K \cite{Schrunk2022}, a temperature too high for quantum oscillations to be observed in this material. 
Therefore, in order to examine our conjecture, we chose two other $RE$Bi compounds with significantly lower $T_N$: DyBi ($T_N=11.2$\,K \cite{Hulliger1980}) and HoBi ($T_N=5.7$\,K \cite{Hulliger1984}), and investigated their magnetotransport properties.
HoBi has previously been reported to exhibit XMR, magnetic-field induced resistivity plateau and complex magnetic structure \cite{Yang2018a,Wu2019d,Fente2013}, which we also confirmed in our study.
In contrast to HoBi, magnetotransport properties of DyBi were not studied. We found that, similar to HoBi, DyBi exhibits XMR and magnetic field-induced resistivity plateau.
In addition, we observed that oscillation frequencies change with temperature variation in both compounds, which corresponds to a change in the topology of Fermi surface. 

\section{Results and discussion}

\begin{figure}[h]
	\centering
	\includegraphics[width=\linewidth]{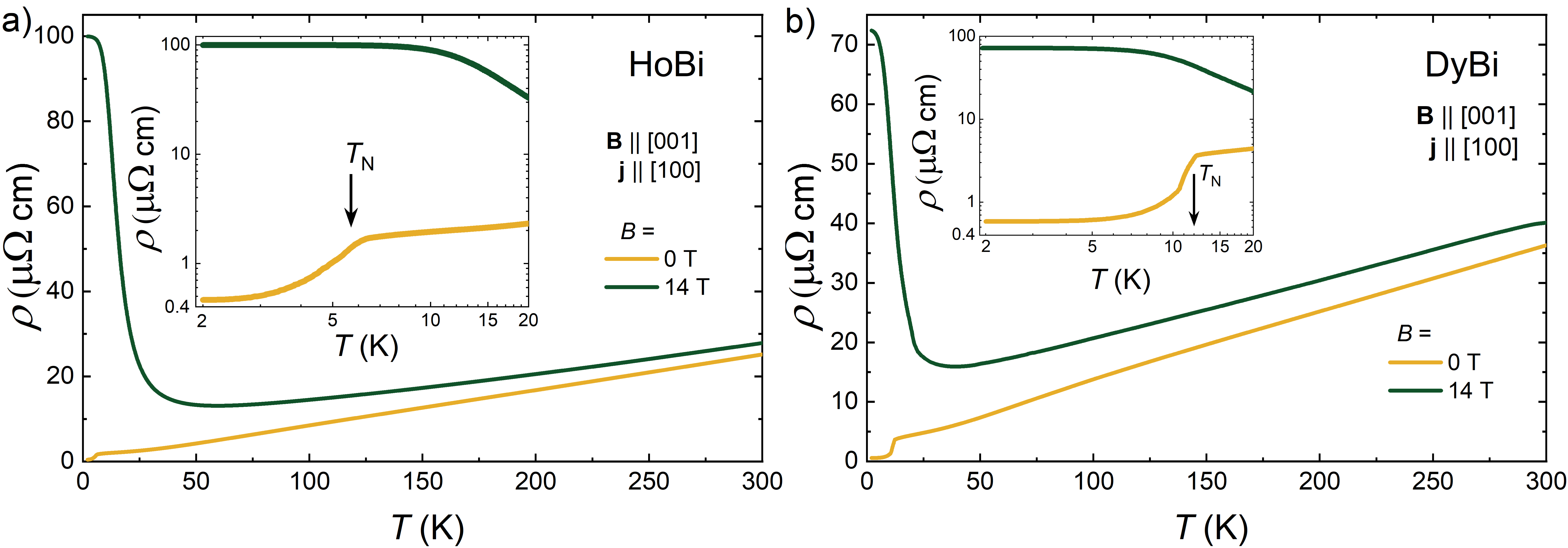}
	\caption{Temperature dependence of the electrical resistivity of HoBi (a) and DyBi (b), measured in zero magnetic field (yellow line) and in magnetic field of 14\,T (green line), applied transverse to the electrical current ({\bf{j}}) direction. 
		Insets show the low-temperature data, with arrows indicating N\'{e}el temperatures.
		\label{Fig2}}
\end{figure}

\subsection{Electrical resistivity}

In zero magnetic field, the temperature dependence of electrical resistivity, $\rho(T)$, of both compounds (see Figure~\ref{Fig2}\,a,b) displays metallic-like behavior, with $\rho$ decreasing as the temperature decreases, with values of $\rho= 25.6$\,$\mu$$\Omega$\,\rm{cm} and $\rho= 40.2$\,$\mu$$\Omega$\,\rm{cm} at $T= 300$\,K for HoBi and DyBi, respectively.
At $T=2$\,K electrical resistivity falls to $\rho= 0.46$\,$\mu$$\Omega$\,\rm{cm} and $\rho= 0.63$\,$\mu$$\Omega$\,\rm{cm}, for HoBi and DyBi, respectively (and the residual resistivity ratios $[\equiv\rho(300\,{\rm K})/\rho(2\,{\rm K})]$ are 56 and 64).
A significant drop in electrical resistivity is observed as the temperature decreases below $T=5.9$\,K for HoBi and $T=12$\,K for DyBi. 
These drops can be attributed to the elimination of spin-disorder scattering by the antiferromagnetic ordering, which according to the literature takes place at $T_N=5.7$\,K and $T_N=11$\,K for HoBi and DyBi, respectively.~\cite{Fente2013,Hulliger1980,Wada1995}
Magnetic field of 14\,T, applied transverse to the current, dramatically changes the $\rho(T)$ dependence: resistivity is enhanced in whole temperature range, and displays metallic-like behavior down to 60\,K and 40\,K,  
for HoBi and DyBi, respectively, followed by a sharp increase of $\rho$ with further decreasing temperature. 
At low temperatures magnetic field-induced resistivity plateaus are observed (see insets to Figure~\ref{Fig2}a,b) for both compounds. 
Similar behavior of $\rho(T)$ dependence has been reported in other rare earth bismuthides and antimonides: LuSb, LuBi, YBi, LaBi, LaSb, YSb, PrBi, GdSb, TbSb
~\cite{Pavlosiuk2017,Pavlosiuk2018,Niu2016,Pavlosiuk2016,Vashist2019, Song2018b,Tang2022} and in several topological semimetals: ZrSiS, NbP, TaAs.
~\cite{Ali2016,Shekhar2015,Sankar2018}
Appearance of the magnetic field-induced resistivity plateau has been attributed to the metal-insulator transition,~\cite{Singha2016a,Li2016f} perfect or nearly perfect carrier compensation\cite{Wang2015d,Xu2017f} or Lifshitz transition.~\cite{Wu2015}
In most of $RE$Sb and $RE$Bi compounds, the overall $\rho(T)$ dependence in applied magnetic field (including resistivity plateau and change of behaviour from metallic- to semiconducting-like) has been described in terms of carrier compensation.~\cite{Pavlosiuk2018,Niu2016,Pavlosiuk2016, Pavlosiuk2017,Kumar2016,Xu2017f,Ghimire2016}
For HoBi, the shape of $\rho(T)$ obtained in $B=14$\,T is similar to that of temperature dependent Hall coefficient ($R_H(T)$) obtained for $B=14$\,T, shown in the inset to Figure~\ref{Fig_Hall}. 

\subsection{Magnetoresistance}

\begin{figure}
	\includegraphics[width=\linewidth]{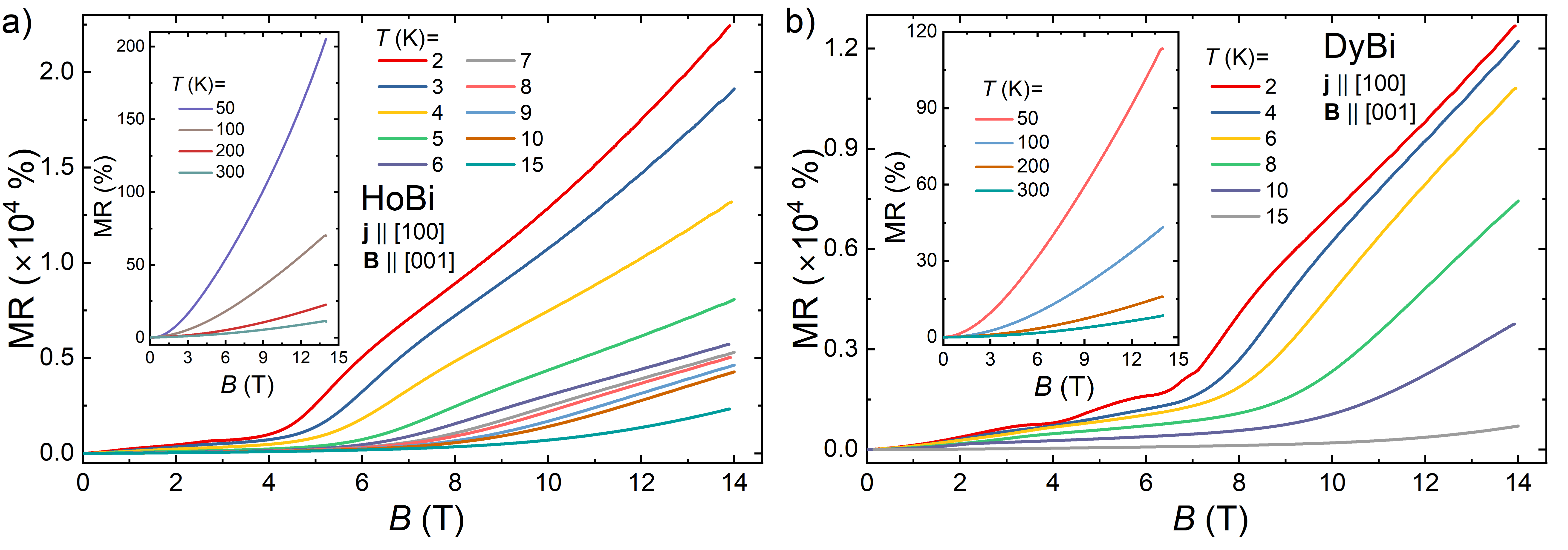}
	\caption{Transverse ({\bf{j$\perp$B}}) magnetoresistance as a function of magnetic field for HoBi (a) and DyBi (b), measured at several temperatures.
		\label{Fig3}}
\end{figure}

Figure~\ref{Fig3} shows magnetoresistance, MR$\,=\,100\%\,\times {[\rho(B)-\rho(0)]}/{\rho(0)}$ as a function of magnetic field applied transverse to the current, measured at several temperatures.
Both compounds exhibit extremely large, positive and non-saturating MR, reaching (at $T=2$\,K, in $B=14$\,T) 2.2\,$\times$\,10$^4$\,\% and 1.3\,$\times$\,$10^4$\,\% for HoBi and DyBi, respectively.
With increasing temperature MR gradually decreases, down to 11.3\,\% and 8.6\,\% for HoBi and DyBi, respectively (at $T=300$\,K and in $B=14$\,T).
The MR values which we obtained for HoBi are slightly smaller than those reported in literature for this material, but the order of magnitude of MR is the same.~\cite{Wu2019d}
In case of DyBi, the magnitude of MR is almost the same as that reported for DySb in Ref.~\cite{Liang2018c}. 
Values of MR in both compounds are related to the residual resistivity ratios in a similar way as in other $REPn$.\cite{Pavlosiuk2018}
XMR exhibited by $REPn$ series has been attributed to: carrier compensation,~\cite{Pavlosiuk2018,Pavlosiuk2017,Pavlosiuk2016,Fan2020,Song2018b,Ghimire2016} metal-insulator transition,~\cite{Tafti2015} or $d\mbox{-}p$ orbital mixing in the presence of carrier compensation.~\cite{Tafti2016a}

\begin{figure}
	\includegraphics[width=\linewidth]{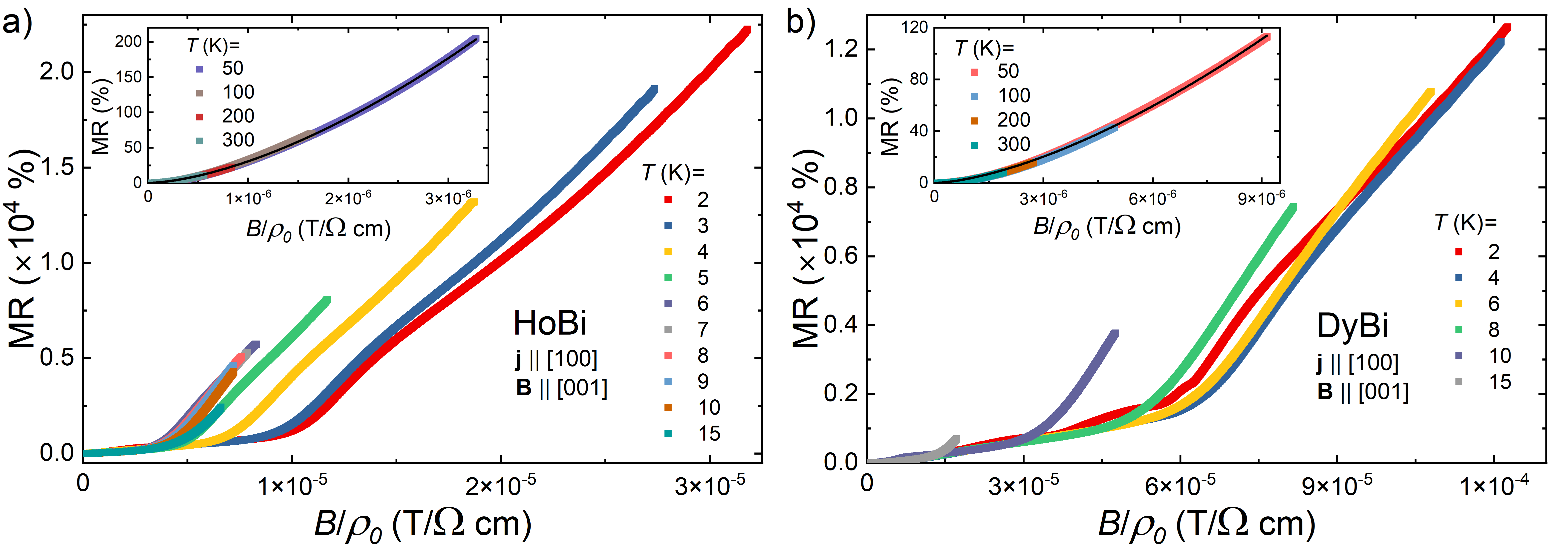}
	\caption{Standard Kohler's scaling for HoBi (a) and DyBi (b). Main panels show the data taken in the temperature range 2-15\,K; insets show the data recorded in the temperature range 50-300\,K. Black solid lines in the insets correspond to the fits with Kohler's formula.  
		\label{Fig4}}
\end{figure}

In numerous, both magnetic and non-magnetic $REPn$ demonstrating XMR, nearly quadratic behavior of MR$\propto B^m$ ($m<2$) was observed.~\cite{Wang2018, Hu2018b, Song2018b, Wu2018c} 
However, for HoBi and DyBi, MR$(B)$ isotherms for temperatures $\leq 15$\,K display strikingly different character.
Those for $T<T_N$ exhibit two distinct regions, 
for example, at $T=2$\,K, in the low field region, MR$(B)$ increases with fluctuating slope up to about 5\,T for HoBi, and up to about 8\,T for DyBi. 
Such behavior can be related to the series of metamagnetic  transitions reported for both materials.~\cite{Fente2013,Yang2018a,Wu2019d, Hulliger1980,Zhao2022}
Magnetic moments of Ho and Dy become fully aligned with the stronger magnetic field, and in such conditions the onset of typical MR ($\propto B^m$) occurs in both compounds.  
Interestingly, our magnetotransport results for HoBi are in a full agreement with those reported by Wu et al.\cite{Wu2019d}, although they differ from those described by Yang et al.\cite{Yang2018a} That difference comes from the magnetic anisotropy in HoBi, since in those measurements direction of the applied magnetic field was different ([001] and [110], respectively).
In Ref.~\cite{Yang2018a} it has been concluded that HoBi is the first material where XMR appears in the fully field-polarized magnetic state. Our results indicate that the mechanism of XMR in DyBi is the same as in HoBi.      

It has recently been proposed that the magnetotransport properties of semimetals displaying XMR can be understood in the scope of Kohler's scaling, which reflects the degree of charge carrier compensation.~\cite{Wang2015} According to standard Kohler's rule $MR=(B/\rho_0)^m$, where $m$ is a sample characteristic parameter, indicating the level of charge balance in the system.
For a perfectly electron-hole compensated material $m=2$. 
In order to test this approach for our materials, we performed the Kohler's scaling, as shown in Figure~\ref{Fig4}. 
Kohler's rule is violated for both compounds from 2 to 15\,K (see main panels of Figure~\ref{Fig4}a,b). i.e. the temperature range covering antiferromagnetic order and noticeable short-range order.
In several compounds exhibiting extreme magnetoresistance, e.g. LaBi, YBi, WP$_2$, WTe$_2$ or TaAs, Kohler’s rule is violated as well, \cite{Sankar2018,Sun2016a,Qian2018,Wang2017b,Wu2015} but in these cases violation is due to the change of  concentration and mobility of carriers with temperature, not to the magnetic interactions. 
On the other hand, in the temperature interval from 50 to 300\,K MR isotherms of HoBi and DyBi collapse onto a single curve (see insets to Figure~\ref{Fig4}a,b). 
For the MR data obtained in the temperature range 50-300\,K, the least-squares fitting (black solid lines in the insets to Figure~\ref{Fig4}a,b) yielded $m=1.6$ and $m=1.54$, for HoBi and DyBi, respectively. 
Similarly, it has recently been shown that Kohler's scaling is effective at $T>50$\,K in HoSb and ErBi.~\cite{Zhang2021c, Fan2020} 
For other $REPn$ containing magnetic $RE$, Kohler's rule was disobeyed in the entire covered range of temperature.~\cite{Tang2022,Wu2019d,Song2018b}
The magnitude of $m$ that we obtained for both compounds is one of the smallest among $REPn$.~\cite{Pavlosiuk2017, Pavlosiuk2018,Pavlosiuk2016, Sun2016a,Fan2020,Kumar2016}
To the best of our knowledge, the only material among $REPn$ compounds with $m$ smaller that we observed in this work is LaAs with $m=1.35$.~\cite{Yang2017c}

\begin{figure}[h]
	\includegraphics[width=0.49\textwidth]{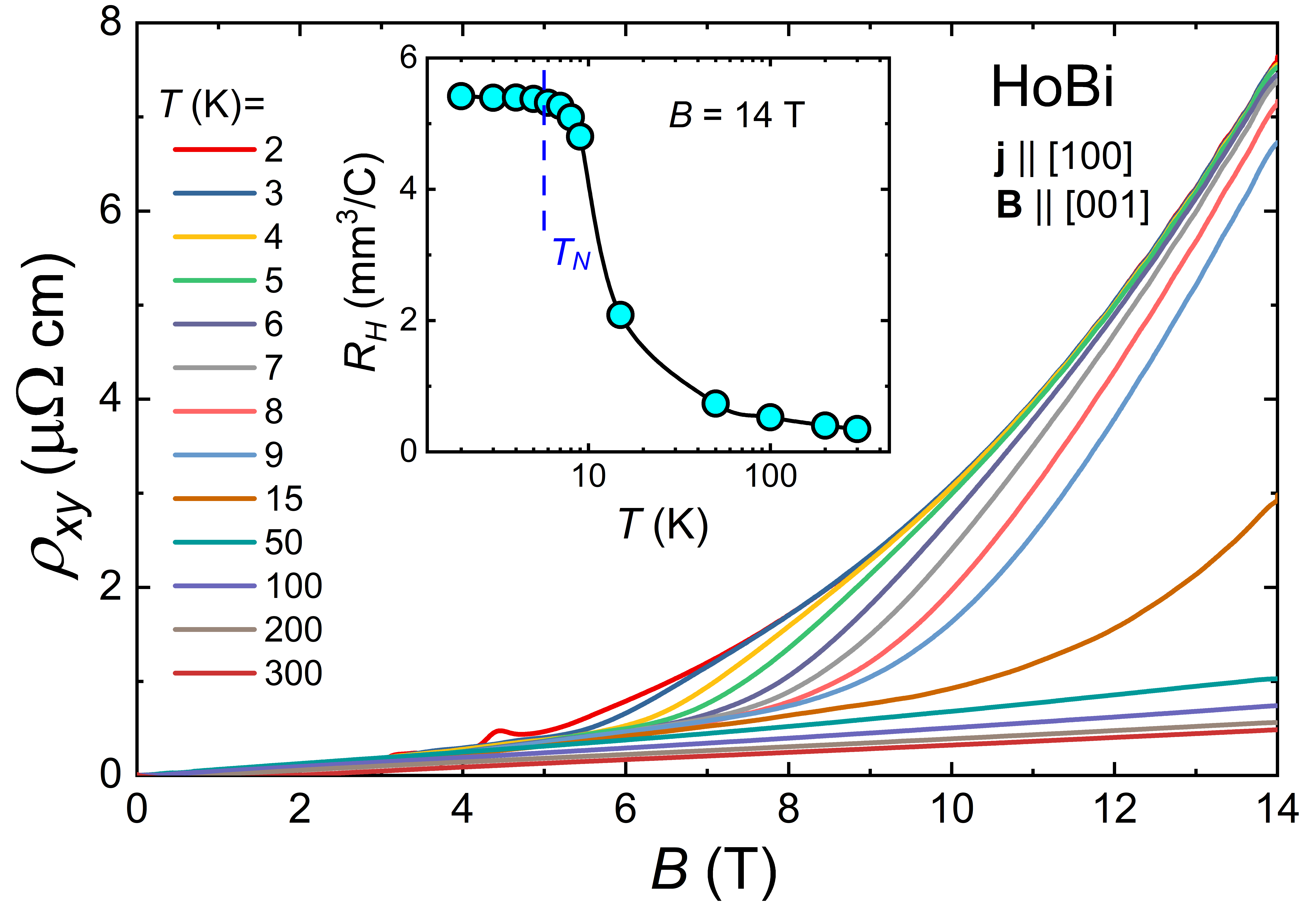}
	\caption{Hall resistivity ($\rho_{xy}$) of HoBi as a function of magnetic field taken at several different temperatures. Inset: temperature dependence of Hall coefficient in $B=14$\,T. Vertical dashed line indicates the N\'{e}el temperature.   
		\label{Fig_Hall}}
\end{figure}

The Kohler's rule is obeyed in simple non-magnetic metals, with one type of carriers and single scattering time \cite{Ziman2001} and in semimetals, only if mobilities of both types of carriers are small, and their ratio, as well as carrier concentrations are all temperature-independent.~\cite{Pavlosiuk2020} 
Therefore, violation of the Kohler's rule for HoBi and DyBi seems most likely due to the temperature-dependent carrier concentrations and/or mobilities.
Results of the Hall effect data presented in Figure~\ref{Fig_Hall} support such assumption. 
The Hall coefficient ($R_H=\rho_{xy}/B=1/(en_H)$), which is inversely proportional to carrier concentration ($n_H$) varies strongly with increasing temperature (see inset to Figure~\ref{Fig_Hall}).
Moreover, the frequency of observed Shubnikov-de Haas (SdH) oscillations also depend on temperature (see below), which is consistent with the assumption of temperature-dependent carrier concentrations in HoBi and DyBi.

\begin{figure}[h]
	\includegraphics[width=\linewidth]{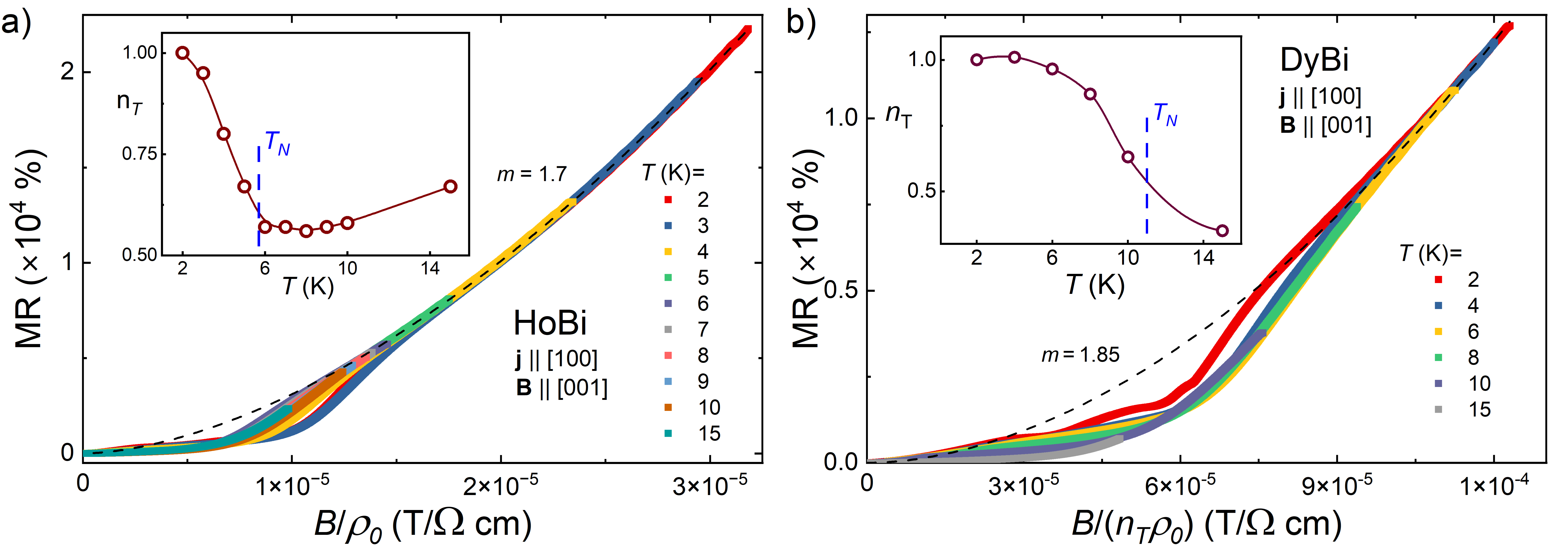}	
	\caption{Extended Kohler's scaling, $MR=(B/(n_T\rho_0))^m$, for HoBi (a) DyBi (b); black dashed lines show power functions representing Kohler's rule. Insets show changes of the factor $n_T$ with temperature (lines are guides for the eye). Blue dashed lines indicate the N\'{e}el temperatures.  
		\label{figExtKohl4}}
\end{figure}

Recently, an extended approach to Kohler's rule has been reported, $MR=(B/(n_T\rho_0))^m$ in which temperature-dependent carrier concentration is implemented through introducing a factor $n_T$.~\cite{Xu2021d}
We tested that approach to our data for temperatures where standard Kohler's scaling failed. We demonstrate the outcome in Figure~\ref{figExtKohl4}. For HoBi extended scaling is working, 
for $T\leq 15$\,K all data above MR$\sim 5000$\% collapse on single curve with $m=1.7$. This exponent is different from 1.6 obtained for $T\geq 50$\,K, which indicates small but clear change of electron-hole compensation between these two temperature ranges. In the case of DyBi extended scaling works, but the curve on which data collapse (for $4\!<\!T\!<\!6$\,K, and MR$\sim 9000$\%) corresponds to $m=1.85$, suggesting a higher degree of charge carriers compensation if compared with $T\geq 50$\,K, for which $m=1.54$. 
Parameter $n_T$ strongly evolves with temperature, for both compounds, as shown in insets to Figure~\ref{figExtKohl4}a,b, implying a significant change of carrier concentration (we will discuss this further below when analyzing the SdH oscillations). 
In this range of $T$, strong interactions of magnetic moments, and series of metamagnetic transitions strongly influence scattering rate of charge carriers, and magnetoresistance regains its typical (power function) behavior only in strong magnetic fields, when antiferromagnetic order is destroyed and critical fluctuations damped. 

\subsection{Shubnikov-de Haas oscillations}

\begin{figure}[h]
	\includegraphics[width=\linewidth]{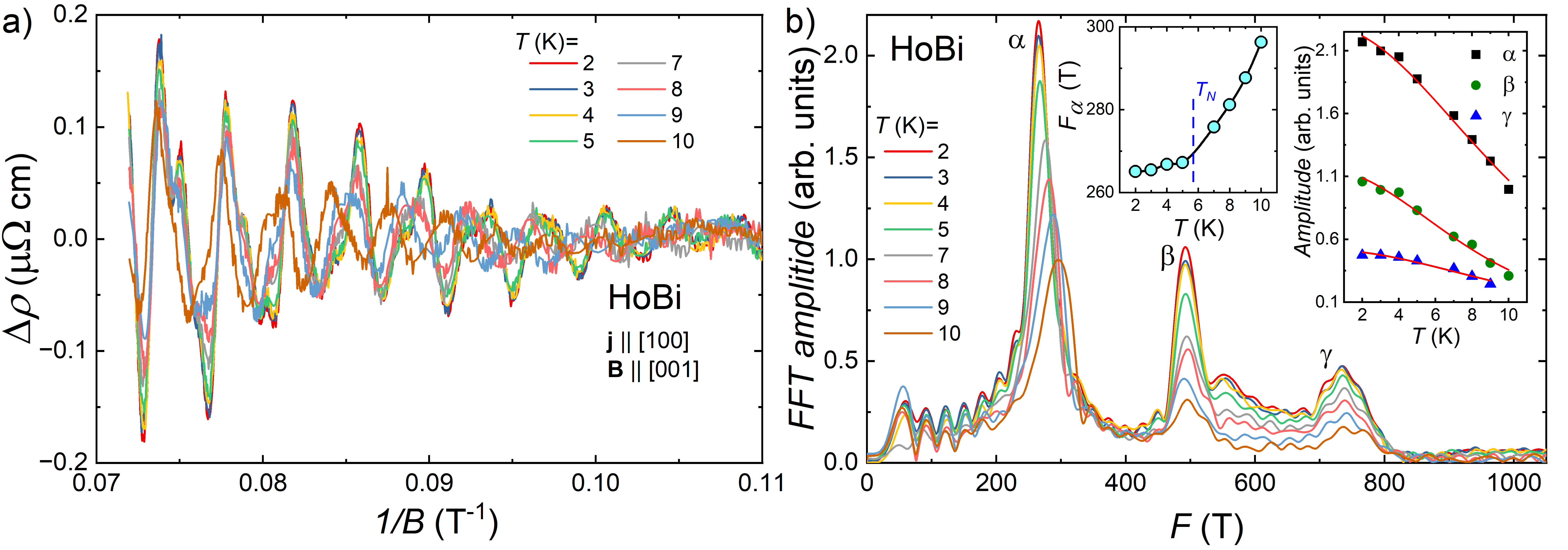}
	\caption{(a) The SdH oscillations extracted from electrical resistivity data in transverse magnetic field at different temperatures for HoBi.
		(b) Fast Fourier transform analysis of the data presented in (a). Left inset: Amplitude of $F_{\alpha}$ principal frequencies in FFT spectra as a function of temperature (black solid line is guide for the eye). Blue dashed line indicates the N\'{e}el temperature. Right inset: Temperature dependence of the FFT peak height. Red solid lines represent the fits of Equation~\ref{LK_eq} to the experimental data.
		\label{Fig6}}
\end{figure}

\begin{figure}[h]
	\includegraphics[width=\linewidth]{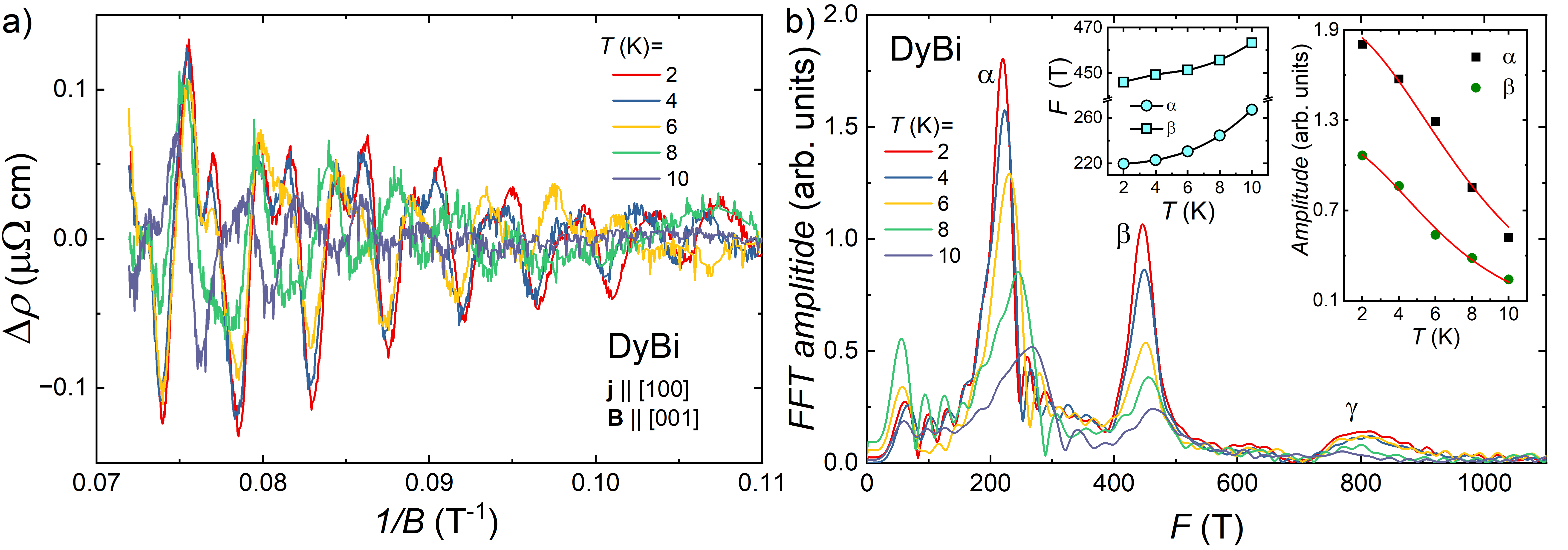}
	\caption{The SdH oscillations extracted from electrical resistivity data in transverse magnetic field at different temperatures for DyBi.
		(b) Fast Fourier transform analysis of the data presented in (a). Left inset: Amplitudes of principal frequencies in FFT spectra as a function of temperature (black solid lines are guides for the eye). Right inset: Temperature dependence of the FFT peak height. Red solid lines represent the fits of Equation~\ref{LK_eq} to the experimental data.
		\label{Fig7}}
\end{figure}

Magnetoresistance isotherms, MR$(B)$, for both compounds show oscillatory behavior above $B=9$\,T, which we ascribe to the SdH effect. 
For both materials, the SdH oscillations survive at quite high temperatures, up to at least 10\,K.
Figures~\ref{Fig6}a, and \ref{Fig7}a depict the oscillating part of the electrical resistivity ($\Delta\rho$) as a function of the inverse magnetic field ($1/B$), obtained by subtraction of a smooth background from the resistivity data.
Fast Fourier transform (FFT) analysis of these oscillations revealed three fundamental frequencies ($F_{\alpha}$, $F_{\beta}$ and $F_{\gamma}$ listed in Table~\ref{Tab_Oscillations}), which points at possible complex form of Fermi pockets or/and at multi-band type of conductivity. 
$F_{i}$ values we obtained for HoBi are very close to those reported in Ref.~\cite{Yang2018a}. 
In contrast to that report, we did not observe the high-frequency oscillations ($F>900$\,T), most probably due to the narrower range of magnetic field in which our experiments were carried out. 
For DyBi, our observation of the SdH oscillations is the first. and the frequencies we obtained (see Table~\ref{Tab_Oscillations}) are quite close to those found for HoBi, which implies that relevant sheets of Fermi surfaces in both compounds are rather similar.
According to the results of several theoretical calculations, Fermi surfaces of $REPn$ compounds, which contain heavy rare earth elements, are qualitatively similar, all of them possess two or three nearly isotropic hole-like pockets in the center of the Brillouin zone and strongly anisotropic triplicate electron-like pocket. \cite{Tang2022,Liang2018c,Wu2019d,Zhang2021c, Xie2020,Pavlosiuk2018,Pavlosiuk2017,Kakihana2019,Hosen2020a}
In the magnetic-field induced polarized state (with all spins aligned with strong enough magnetic field) the electronic structure of magnetic $REPn$ changes with each Fermi pocket splitting into two pockets due to the magnetic-field induced time-reversal symmetry breaking. \cite{Zhang2021c,Wu2019d,Tang2022,Yang2018a} 
Such splitting has already been reported for HoBi.~\cite{Wu2019d,Yang2018a}
Comparing the frequencies we obtained to those reported in Ref.\,\cite{Yang2018a}, we may assume that $F_{\alpha}$ originates from the smallest hole-like Fermi pocket (denoted as $h_1$ in Ref.\,\cite{Yang2018a}); $F_{\beta}$ and $F_{\gamma}$ come from two extreme cross-sections of the electron-like pocket (denoted as $e_2$ in Ref.\,\cite{Yang2018a}).
For another $REPn$, DySb, quantum oscillations were also reported \cite{Liang2018c} and their frequencies are listed in Table~\ref{Tab_Oscillations} along those theoretically calculated. 
The authors of Ref.~\cite{Liang2018c} ascribed the frequencies $F_{\alpha}$ and $F_{\gamma}$ to the electron-like Fermi pocket and frequency $F_{\beta}$ to the hole-like pocket. 
This assignment is different from that made for HoBi in Ref.~\cite{Yang2018a}
To find which option is valid for HoBi and DyBi, it might be necessary to perform angle-dependent studies of quantum oscillations and to calculate electronic structures for both materials.

According to the Onsager relation, the oscillation frequencies are directly proportional to the area of the extreme cross-sections ($S_i$) of Fermi surface pockets: 
$F_i=(hS_i)/e$, where $h$ stands for the Planck constant, and $e$ is the elementary charge \cite{Shoenberg1984}. 
Assuming, that the shapes of hole-like and electron-like Fermi pockets of both compounds can be approximated by a sphere and an ellipsoid, respectively, we calculated the corresponding Fermi wave vectors ($k_F$) and charge carrier concentrations $(n=V_F/(4\pi^3$), where $V_F$ is volume of the Fermi pocket). 
The calculated values (listed in Table~\ref{Tab_Oscillations}) are smaller than those obtained from our Hall effect results, as well as reported in literature.~\cite{Wu2019d}
This difference is most likely due to the fact that not all Fermi pockets were detected with the SdH oscillations.    

Interestingly, in the FFT spectra for both compounds, clear shifts of the peaks related to the $F_{\alpha}$ were observed. 
In the case of DyBi, the pronounced shift was also noticed for $F_{\beta}$. 
To see it more clearly, the temperature dependence of $F_i$ are shown in the left insets to Figure~\ref{Fig6}b and Figure~\ref{Fig7}b, where one can see that $F_i$ increases with increasing temperature.
Similar behavior has previously been noticed for several $REPn$, such as DySb,~\cite{Liang2018c} SmBi,~\cite{Sakhya2021a} HoSb,~\cite{Wang2018b} and NdSb.~\cite{Wang2018a}     
In the first three reports, the authors did not comment on these observations, while in the fourth one concluded that it indicated the change of topology of the Fermi surface. 
Interestingly, for $REPn$ containing non-magnetic $RE$, there were no reports on the temperature dependence of oscillations frequencies, which may point that magnetism is responsible for this behavior. 
From another point of view, it is also known that in semimetals with small Fermi pockets and small effective masses, the Fermi level can be shifted with increasing temperature, leading to changes in the Fermi surface.~\cite{Shoenberg1984}
For our materials, the Fermi temperature ($T_F=E_F/k_B$) is much larger than temperatures at which SdH oscillations were observed, therefore we can exclude the above scenario.
The thermal expansion of the material is another possible source of the Fermi level shift.~\cite{Shoenberg1984} 
In antiferromagnets, at temperatures close to the $T_N$ thermal expansion very often demonstrates drastic change,~\cite{Pavlosiuk2018c,Budko2008} which was observed for HoBi as well.~\cite{Hulliger1984}         
However, thermal expansion as the only source of change of the Fermi surface can also be ruled out for the following reasons: (i) for HoBi above $T_N$ thermal expansion does not change significantly, but $F_{\alpha}$ continues to increase; (ii) for DyBi both $F_{\alpha}$ and $F_{\beta}$ increase with increasing temperature, but if a simple Fermi level shift would be the reason of this, then one of these frequencies should decrease and the other increase with increasing temperature, since we ascribe them to Fermi pockets of different type, electron- and hole-like.
Very recently, it has been proposed that change of oscillation frequency can be a signature of Dirac bands \cite{Guo2021a}, however, in this case magnitude of frequency change is two orders of magnitude smaller than we observed for DyBi and HoBi. 
It has recently been shown that in several $REPn$ compounds a new type of magnetic splitting leads to the appearance of additional Fermi arcs below $T_N$,~\cite{Schrunk2022,Kushnirenko2022} moreover, such splitting can change the shape of the 3D Fermi surface. 
Our analysis of the SdH oscillations did not reveal new oscillation frequencies due the Fermi arcs, but we can speculate that the strong temperature dependence of oscillation frequencies comes from a new type of magnetic splitting, especially because the rate of change of $F$ with temperature is very different below and above $T_N$ (cf. left inset to Figure~\ref{Fig6}b). That splitting is also likely to underlay dramatic change in Kohler's scaling between paramagnetic and antiferromagnetic state of HoBi (cf. inset to Figure~\ref{figExtKohl4}a). 

Recently, temperature variations of the oscillation frequency have been found in MnBi$_{2-x}$Sb$_x$Te$_4$, PrAlNi and CePtBi.~\cite{Jiang2021b,Lyu2020a,Goll2006} 
For MnBi$_{2-x}$Sb$_x$Te$_4$ this variation was attributed to the change in the field-induced magnetization \cite{Jiang2021b}, and for PrAlNi it has been proposed that spin polarization is responsible for this effect.~\cite{Lyu2020a}
We suppose, that the former explanation may also be reasonable for HoBi and DyBi.
In turn, for CePtBi, this rare behavior was attributed to the temperature variation of $4f$ electrons hybridization.~\cite{Goll2006}  
This explanation can be excluded in our case, as the electronic bands related to the $4f$ electrons are located much above the Fermi level in $REPn$.~\cite{Tang2022}

\begin{table}
	\caption{Parameters obtained from the analysis of the Shubnikov-de Haas oscillations for HoBi and DyBi, compared to the data reported for HoBi in Ref.~\cite{Yang2018a} and for DySb in Ref.~\cite{Liang2018c}.}
	\centering
	\begin{tabular*}{0.8\textwidth}{@{\extracolsep{\fill}}*{5}{c}} \hline\hline
		Compound &$i=$ & $\alpha$& ${\beta}$& ${\gamma}$ \\\hline
		HoBi & $F_i$ (T) & 265 & 492 & 735 \\
		(this work)& $k_{F,i}$ (\r{A})& 0.09 & 0.12 & 0.18 \\
		& $n_i$ (cm$^{-3}$) & 2.44$\times10^{19}$&
		\multicolumn{2}{c}{9.22$\times10^{19}$}\\
		& $m^*_i$ ($m_e$) & 0.17 & 0.22 & 0.17\\\hline
		HoBi (Ref.~\cite{Yang2018a}) & $F_{i,calc}$ (T) & 243 & 489 & 835 \\
		DFT calculations& $m^*_{i,calc}$ ($m_e$)& 1.14 & 0.54 & 0.46\\
		\\
		HoBi (Ref.~\cite{Yang2018a}) & $F_i$ (T) & 280 & 513 & 747 \\
		SdH oscillations& $m^*_i$ ($m_e$) & 0.27 & 0.29 & 0.29\\
		\hline\hline
		DyBi &$F_i$ (T) & 220 & 446& 815 \\
		(this work)& $k_{F,i}$ (\r{A})& 0.08 & 0.012 & 0.21 \\
		& $n_i$ (cm$^{-3}$) & 1.85$\times10^{19}$&
		\multicolumn{2}{c}{9.74$\times10^{19}$}\\
		& $m^*_i$ ($m_e$) & 0.22& 0.27& - \\ \hline
		DySb (Ref.~\cite{Liang2018c}) & $F_{i,calc}$ (T) & 369 & 1022 & 1063 \\
		DFT calculations& $m^*_{i,calc}$ ($m_e$)& 0.21 & 0.42 & 0.49\\
		\\
		DySb (Ref.~\cite{Liang2018c}) & $F_i$ (T) & 369 & 745 & 1094 \\
		SdH oscillations& $m^*_i$ ($m_e$) & 0.69 & 0.88 & -\\
		\hline\hline
	\end{tabular*}\label{osc}
	\label{Tab_Oscillations}
\end{table}

In Ref.~\cite{Goll2006}, the authors explained the possible role of the so-called background magnetization and its temperature dependence on the oscillation frequency. 
For CeBiPt described in that paper, this effect has been rejected (because of opposite than expected change of $F$ vs. $T$), but for our results it could be applicable. 
However, such mechanism would affect all observed frequencies in a similar way, whereas for HoBi the value of $F_{\beta}$ does not change at all, in contrast to strong change of $F_{\alpha}$. 
Moreover, in high fields, between 9 and 14\,T, the magnetization is very close to saturation and weakly changes with temperature. These allow us to rule out the role of background magnetization.        

Despite the fact that the Fermi surfaces in both compounds vary with temperature, we made a working assumption that the effective masses are almost temperature independent. 
This allowed us to use the Lifshitz-Kosevich theory for determining the effective masses.~\cite{Shoenberg1984} 
FFT peaks' amplitudes as a function of temperature are shown in the right insets to Figure~\ref{Fig6}b and Figure~\ref{Fig7}b for HoBi and DyBi, respectively. 
Effective masses ($m^*_i$) were obtained through least-square fitting with the equation representing thermal damping of SdH oscillations:
\begin{equation}
	R_i(T)=(\lambda m^*_iT/B_{\rm{eff}})/\sinh(\lambda m^*_iT/B_{\rm{eff}}), (i=\alpha, \beta)
	\label{LK_eq}
\end{equation}
where $B_{\rm{eff}}=2/(1/B_1 + 1/B_2)=10.96$\,T and $\lambda$=14.7\,T/K.
The value of $B_{\rm{eff}}$ was obtained for $B_1=14$\,T and $B_2=9$\,T, limiting magnetic field range of the analysis.
Obtained effective masses for HoBi and DyBi are listed in Table\,\ref{Tab_Oscillations}, these values are similar to those reported for other $REPn$.~\cite{Yang2018a,Pavlosiuk2017,Wang2018b} 

\section{Conclusions}
We have performed a comprehensive study of magnetotransport properties of two XMR materials, DyBi and HoBi. 
While the magnetotransport properties that we observed for HoBi are in agreement with previous reports, \cite{Yang2018a,Wu2019d} no systematic studies of magnetotransport of DyBi have previously been reported. 
For both materials, we observed XMR of the order of $10^4\%$ and magnetic field-induced resistivity plateau at low temperatures. 
We found that standard Kohler's scaling is obeyed at temperature range from 50 to 300\,K. 
On the other hand, satisfying the extended Kohler's rule at $T\leq 15$\,K supports the variation of carrier concentration with temperature.
The results of Hall effect measurements and analysis of the SdH oscillations also confirmed the temperature dependent carrier concentrations. 
The rare change of quantum oscillation frequencies with temperature increasing was observed for both materials and could be attributed to the change of the Fermi surface shape due to the emergence of the new type of Fermi arcs below $T_N$, recently discovered for several $REPn$\cite{Schrunk2022,Kushnirenko2022}. 
This mechanism is strongly supported by the clear difference in both, $n_T(T)$ and $F(T)$ behaviour at temperatures below and above $T_N$.
Similarly to other $REPn$, the SdH oscillations in both studied compounds show multi-frequency character, where one of the frequencies comes from the hole-type Fermi pocket and others from electron-type Fermi pocket. 
This suggests that electron-hole compensation can be responsible for XMR in these compounds.\\
\newpage
\textbf{METHODS}

\textbf{Sample synthesis and characterization}

High quality single crystals of $RE$Bi ($RE$=\,Ho,\,Dy) were synthesized by the self-flux growth method with the molar ratio of constituent elements $RE$ (purity of 99.9 wt.\%)\,:\,Bi (purity of 99.9999 wt.\%)\,=\,1\,:\,20. 
The charge was heated in sealed quartz ampule up to 1373\,K, held for 24\,h at this temperature and then cooled to 873\,K with a step of 2\,K/h. 
The excess of Bi flux was removed using a centrifuge. 
The largest among obtained single crystals had a size of $3\times3\times3$\,mm$^3$.
The $RE$Bi single crystals oxidize easily in ambient atmosphere and decompose after several hours. 
Samples were examined at room temperature using X-ray powder diffraction (XPRD) technique on the X'pert Pro (PANalytical) diffractometer with Cu-K$_\alpha$ radiation. 
The crystal structure refinement (by the Rietveld method) was performed using Fullprof software.~\cite{Rodriguez-Carvajal1993}
The obtained lattice parameters ($a=6.225$\,{\AA} and $a=6.249$\,{\AA} for HoBi and DyBi, respectively) are in a good agreement with data reported in literature.~\cite{Wu2019d,Yoshihara1975}
The single crystals were oriented, and their quality was checked by the Laue method using the Laue-COS (Proto) system. 

\textbf{Electrical transport measurements}

Electrical transport properties, including electrical resistivity, Hall effect and magnetoresistance, were measured using a four-probe method employing the Physical Property Measurement System (PPMS, Quantum Design). Measurements were carried out in the temperature range 2-300\,K and in the magnetic fields up to 14\,T, with ac electrical current flowing along [100] crystallographic direction.\\

\textbf{DATA AVAILABILITY} 

The data that support the findings of this study are available from the corresponding author upon reasonable request.\\

\textbf{Acknowledgements}

This work was supported by the National Science Centre (Poland) under research grant 2021/40/Q/ST5/00066. 
For the purpose of Open Access, the author has applied a CC-BY public copyright license to any Author Accepted Manuscript (AAM) version arising from this submission.
O.P. is thankful to the Max Planck Institute for Chemical Physics of Solids for their hospitality and support during a part of the work on this project. 
We thank K.\,Dyk for performing powder X-ray diffractography.\\

\textbf{AUTHOR CONTRIBUTIONS} 

D.K., P.W. and O.P. conceived the study; O.P. and P.N. conducted all the experiments; O.P., P.N. and P.W. analysed the results; O.P. and P.N. wrote the manuscript with inputs from all authors. 
All authors discussed the results and edited the manuscript.\\


\begin{thebibliography}{10}
	\expandafter\ifx\csname url\endcsname\relax
	\def\url#1{\texttt{#1}}\fi
	\expandafter\ifx\csname urlprefix\endcsname\relax\def\urlprefix{URL }\fi
	\providecommand{\bibinfo}[2]{#2}
	\providecommand{\eprint}[2][]{\url{#2}}
	
	\bibitem{Zide2006}
	\bibinfo{author}{Zide, J.~M.} \emph{et~al.}
	\newblock \bibinfo{title}{Increased efficiency in multijunction solar cells
		through the incorporation of semimetallic {E}r{A}s nanoparticles into the
		tunnel junction}.
	\newblock \emph{\bibinfo{journal}{Appl. Phys. Lett.}}
	\textbf{\bibinfo{volume}{88}}, \bibinfo{pages}{162103}
	(\bibinfo{year}{2006}).
	
	\bibitem{LIU201156}
	\bibinfo{author}{Liu, X.} \emph{et~al.}
	\newblock \bibinfo{title}{{Properties of molecular beam epitaxially grown
			ScAs:InGaAs and ErAs:InGaAs nanocomposites for thermoelectric applications}}.
	\newblock \emph{\bibinfo{journal}{J. Cryst. Growth}}
	\textbf{\bibinfo{volume}{316}}, \bibinfo{pages}{56--59}
	(\bibinfo{year}{2011}).
	
	\bibitem{Krivoy2018}
	\bibinfo{author}{Krivoy, E.~M.} \emph{et~al.}
	\newblock \bibinfo{title}{{Rare-Earth Monopnictide Alloys for Tunable,
			Epitaxial, Designer Plasmonics}}.
	\newblock \emph{\bibinfo{journal}{ACS Photonics}} \textbf{\bibinfo{volume}{5}},
	\bibinfo{pages}{3051--3056} (\bibinfo{year}{2018}).
	
	\bibitem{Schrunk2022}
	\bibinfo{author}{Schrunk, B.} \emph{et~al.}
	\newblock \bibinfo{title}{{Emergence of Fermi arcs due to magnetic splitting in
			an antiferromagnet}}.
	\newblock \emph{\bibinfo{journal}{Nature}} \textbf{\bibinfo{volume}{603}},
	\bibinfo{pages}{610--615} (\bibinfo{year}{2022}).
	
	\bibitem{Inbar2022}
	\bibinfo{author}{Inbar, H.~S.} \emph{et~al.}
	\newblock \bibinfo{title}{{Epitaxial growth, magnetoresistance, and electronic
			band structure of GdSb magnetic semimetal films}}.
	\newblock \emph{\bibinfo{journal}{Phys. Rev. Materials}}
	\textbf{\bibinfo{volume}{6}}, \bibinfo{pages}{L121201}
	(\bibinfo{year}{2022}).
	
	\bibitem{Chatterjee2019a}
	\bibinfo{author}{Chatterjee, S.} \emph{et~al.}
	\newblock \bibinfo{title}{{Weak antilocalization in quasi-two-dimensional
			electronic states of epitaxial LuSb thin films}}.
	\newblock \emph{\bibinfo{journal}{Phys. Rev. B}} \textbf{\bibinfo{volume}{99}},
	\bibinfo{pages}{125134} (\bibinfo{year}{2019}).
	
	\bibitem{Tafti2015}
	\bibinfo{author}{Tafti, F.~F.}, \bibinfo{author}{Gibson, Q.~D.},
	\bibinfo{author}{Kushwaha, S.~K.}, \bibinfo{author}{Haldolaarachchige, N.} \&
	\bibinfo{author}{Cava, R.~J.}
	\newblock \bibinfo{title}{{Resistivity plateau and extreme magnetoresistance in
			LaSb}}.
	\newblock \emph{\bibinfo{journal}{Nat. Phys.}} \textbf{\bibinfo{volume}{12}},
	\bibinfo{pages}{272} (\bibinfo{year}{2016}).
	
	\bibitem{Pavlosiuk2017}
	\bibinfo{author}{Pavlosiuk, O.}, \bibinfo{author}{Kleinert, M.},
	\bibinfo{author}{Swatek, P.}, \bibinfo{author}{Kaczorowski, D.} \&
	\bibinfo{author}{Wiśniewski, P.}
	\newblock \bibinfo{title}{{Fermi surface topology and magnetotransport in
			semimetallic LuSb}}.
	\newblock \emph{\bibinfo{journal}{Sci. Rep.}} \textbf{\bibinfo{volume}{7}},
	\bibinfo{pages}{12822} (\bibinfo{year}{2017}).
	
	\bibitem{Pavlosiuk2016}
	\bibinfo{author}{Pavlosiuk, O.}, \bibinfo{author}{Swatek, P.} \&
	\bibinfo{author}{Wiśniewski, P.}
	\newblock \bibinfo{title}{{Giant magnetoresistance, three-dimensional Fermi
			surface and origin of resistivity plateau in YSb semimetal}}.
	\newblock \emph{\bibinfo{journal}{Sci. Rep.}} \textbf{\bibinfo{volume}{6}},
	\bibinfo{pages}{38691} (\bibinfo{year}{2016}).
	
	\bibitem{Niu2016}
	\bibinfo{author}{Niu, X.~H.} \emph{et~al.}
	\newblock \bibinfo{title}{{Presence of exotic electronic surface states in LaBi
			and LaSb}}.
	\newblock \emph{\bibinfo{journal}{Phys. Rev. B}} \textbf{\bibinfo{volume}{94}},
	\bibinfo{pages}{165163} (\bibinfo{year}{2016}).
	
	\bibitem{Vashist2019}
	\bibinfo{author}{Vashist, A.}, \bibinfo{author}{Gopal, R.~K.},
	\bibinfo{author}{Srivastava, D.}, \bibinfo{author}{Karppinen, M.} \&
	\bibinfo{author}{Singh, Y.}
	\newblock \bibinfo{title}{{Fermi surface topology and large magnetoresistance
			in the topological semimetal candidate PrBi}}.
	\newblock \emph{\bibinfo{journal}{Phys. Rev. B}} \textbf{\bibinfo{volume}{99}},
	\bibinfo{pages}{245131} (\bibinfo{year}{2019}).
	
	\bibitem{Song2018b}
	\bibinfo{author}{Song, J.~J.} \emph{et~al.}
	\newblock \bibinfo{title}{{Extremely large magnetoresistance in the
			antiferromagnetic semimetal GdSb}}.
	\newblock \emph{\bibinfo{journal}{J. Mater. Chem. C}}
	\textbf{\bibinfo{volume}{6}}, \bibinfo{pages}{3026--3033}
	(\bibinfo{year}{2018}).
	
	\bibitem{Ghimire2016}
	\bibinfo{author}{Ghimire, N.~J.}, \bibinfo{author}{Botana, A.~S.},
	\bibinfo{author}{Phelan, D.}, \bibinfo{author}{Zheng, H.} \&
	\bibinfo{author}{Mitchell, J.~F.}
	\newblock \emph{\bibinfo{journal}{J. Phys. Condens. Matter}}
	\textbf{\bibinfo{volume}{28}}, \bibinfo{pages}{235601}
	(\bibinfo{year}{2016}).
	
	\bibitem{Kumar2016}
	\bibinfo{author}{Kumar, N.} \emph{et~al.}
	\newblock \bibinfo{title}{{Observation of pseudo-two-dimensional electron
			transport in the rock salt-type topological semimetal LaBi}}.
	\newblock \emph{\bibinfo{journal}{Phys. Rev. B}} \textbf{\bibinfo{volume}{93}},
	\bibinfo{pages}{241106} (\bibinfo{year}{2016}).
	
	\bibitem{Wu2019}
	\bibinfo{author}{Wu, Z.} \emph{et~al.}
	\newblock \bibinfo{title}{{Probing the origin of extreme magnetoresistance in
			Pr/Sm mono-antimonides/bismuthides}}.
	\newblock \emph{\bibinfo{journal}{Phys. Rev. B}} \textbf{\bibinfo{volume}{99}},
	\bibinfo{pages}{035158} (\bibinfo{year}{2019}).
	
	\bibitem{Pavlosiuk2018}
	\bibinfo{author}{Pavlosiuk, O.}, \bibinfo{author}{Swatek, P.},
	\bibinfo{author}{Kaczorowski, D.} \& \bibinfo{author}{Wiśniewski, P.}
	\newblock \bibinfo{title}{{Magnetoresistance in LuBi and YBi semimetals due to
			nearly perfect carrier compensation}}.
	\newblock \emph{\bibinfo{journal}{Phys. Rev. B}} \textbf{\bibinfo{volume}{97}},
	\bibinfo{pages}{235132} (\bibinfo{year}{2018}).
	
	\bibitem{Ye2018}
	\bibinfo{author}{Ye, L.}, \bibinfo{author}{Suzuki, T.},
	\bibinfo{author}{Wicker, C.~R.} \& \bibinfo{author}{Checkelsky, J.~G.}
	\newblock \bibinfo{title}{{Extreme magnetoresistance in magnetic rare-earth
			monopnictides}}.
	\newblock \emph{\bibinfo{journal}{Phys. Rev. B}} \textbf{\bibinfo{volume}{97}},
	\bibinfo{pages}{081108} (\bibinfo{year}{2018}).
	
	\bibitem{Feng2018}
	\bibinfo{author}{Feng, B.} \emph{et~al.}
	\newblock \bibinfo{title}{{Experimental observation of node-line-like surface
			states in LaBi}}.
	\newblock \emph{\bibinfo{journal}{Phys. Rev. B}} \textbf{\bibinfo{volume}{97}},
	\bibinfo{pages}{155153} (\bibinfo{year}{2018}).
	
	\bibitem{Jiang2018}
	\bibinfo{author}{Jiang, J.} \emph{et~al.}
	\newblock \bibinfo{title}{{Observation of topological surface states and strong
			electron/hole imbalance in extreme magnetoresistance compound LaBi}}.
	\newblock \emph{\bibinfo{journal}{Phys. Rev. Materials}}
	\textbf{\bibinfo{volume}{2}}, \bibinfo{pages}{024201} (\bibinfo{year}{2018}).
	
	\bibitem{Lou2016}
	\bibinfo{author}{Lou, R.} \emph{et~al.}
	\newblock \bibinfo{title}{{Evidence of topological insulator state in the
			semimetal LaBi}}.
	\newblock \emph{\bibinfo{journal}{Phys. Rev. B}} \textbf{\bibinfo{volume}{95}},
	\bibinfo{pages}{115140} (\bibinfo{year}{2017}).
	
	\bibitem{Nayak2016}
	\bibinfo{author}{Nayak, J.} \emph{et~al.}
	\newblock \bibinfo{title}{{Multiple Dirac cones at the surface of the
			topological metal LaBi}}.
	\newblock \emph{\bibinfo{journal}{Nat. Commun.}} \textbf{\bibinfo{volume}{8}},
	\bibinfo{pages}{13942} (\bibinfo{year}{2017}).
	
	\bibitem{Xu2019}
	\bibinfo{author}{Xu, C.~Q.} \emph{et~al.}
	\newblock \bibinfo{title}{{Extreme magnetoresistance and pressure-induced
			superconductivity in the topological semimetal candidate YBi}}.
	\newblock \emph{\bibinfo{journal}{Phys. Rev. B}} \textbf{\bibinfo{volume}{99}},
	\bibinfo{pages}{024110} (\bibinfo{year}{2019}).
	
	\bibitem{Tafti2016}
	\bibinfo{author}{Tafti, F.~F.} \emph{et~al.}
	\newblock \bibinfo{title}{{Tuning the electronic and the crystalline structure
			of LaBi by pressure: From extreme magnetoresistance to superconductivity}}.
	\newblock \emph{\bibinfo{journal}{Phys. Rev. B}} \textbf{\bibinfo{volume}{95}},
	\bibinfo{pages}{014507} (\bibinfo{year}{2017}).
	
	\bibitem{He2020}
	\bibinfo{author}{He, X.} \emph{et~al.}
	\newblock \bibinfo{title}{{PrBi: Topology meets quadrupolar degrees of
			freedom}}.
	\newblock \emph{\bibinfo{journal}{Phys. Rev. B}}
	\textbf{\bibinfo{volume}{101}}, \bibinfo{pages}{075106}
	(\bibinfo{year}{2020}).
	
	\bibitem{Zhang2021c}
	\bibinfo{author}{Zhang, J.-M.} \emph{et~al.}
	\newblock \bibinfo{title}{{Topological quantum phase transition in the magnetic
			semimetal HoSb}}.
	\newblock \emph{\bibinfo{journal}{J. Mater. Chem. C}}
	\textbf{\bibinfo{volume}{9}}, \bibinfo{pages}{6996--7004}
	(\bibinfo{year}{2021}).
	
	\bibitem{Liang2018c}
	\bibinfo{author}{Liang, D.~D.} \emph{et~al.}
	\newblock \bibinfo{title}{{Extreme magnetoresistance and Shubnikov-de Haas
			oscillations in ferromagnetic DySb}}.
	\newblock \emph{\bibinfo{journal}{APL Mater.}} \textbf{\bibinfo{volume}{6}},
	\bibinfo{pages}{086105} (\bibinfo{year}{2018}).
	
	\bibitem{Tang2022}
	\bibinfo{author}{Tang, F.} \emph{et~al.}
	\newblock \bibinfo{title}{{Anisotropic large magnetoresistance and Fermi
			surface topology of terbium monoantimonide}}.
	\newblock \emph{\bibinfo{journal}{Mater. Today Phys.}}
	\textbf{\bibinfo{volume}{24}}, \bibinfo{pages}{100657}
	(\bibinfo{year}{2022}).
	
	\bibitem{Duan2018}
	\bibinfo{author}{Duan, X.} \emph{et~al.}
	\newblock \bibinfo{title}{{Tunable electronic structure and topological
			properties of LnPn (Ln=Ce, Pr, Sm, Gd, Yb; Pn=Sb, Bi)}}.
	\newblock \emph{\bibinfo{journal}{Commun. Phys.}} \textbf{\bibinfo{volume}{1}},
	\bibinfo{pages}{71} (\bibinfo{year}{2018}).
	
	\bibitem{Kushnirenko2022}
	\bibinfo{author}{Kushnirenko, Y.} \emph{et~al.}
	\newblock \bibinfo{title}{{Rare-earth monopnictides: Family of antiferromagnets
			hosting magnetic Fermi arcs}}.
	\newblock \emph{\bibinfo{journal}{Phys. Rev. B}}
	\textbf{\bibinfo{volume}{106}}, \bibinfo{pages}{115112}
	(\bibinfo{year}{2022}).
	
	\bibitem{Shoenberg1984}
	\bibinfo{author}{Shoenberg, D.}
	\newblock \emph{\bibinfo{title}{Magnetic Oscillations in Metals}}
	(\bibinfo{publisher}{Cambridge University Press},
	\bibinfo{address}{Cambridge}, \bibinfo{year}{1984}).
	
	\bibitem{Hulliger1980}
	\bibinfo{author}{Hulliger, F.}
	\newblock \bibinfo{title}{{Magnetic behavior of DyBi}}.
	\newblock \emph{\bibinfo{journal}{J. Magn. Magn. Mater.}}
	\textbf{\bibinfo{volume}{15-18}}, \bibinfo{pages}{1243--1244}
	(\bibinfo{year}{1980}).
	
	\bibitem{Hulliger1984}
	\bibinfo{author}{Hulliger, F.}, \bibinfo{author}{Ott, H.~R.} \&
	\bibinfo{author}{Siegrist, T.}
	\newblock \bibinfo{title}{{Low temperature behaviour of HoBi}}.
	\newblock \emph{\bibinfo{journal}{J. Less-Common Met.}}
	\textbf{\bibinfo{volume}{96}}, \bibinfo{pages}{263--268}
	(\bibinfo{year}{1984}).
	
	\bibitem{Yang2018a}
	\bibinfo{author}{Yang, H.~Y.} \emph{et~al.}
	\newblock \bibinfo{title}{{Interplay of magnetism and transport in HoBi}}.
	\newblock \emph{\bibinfo{journal}{Phys. Rev. B}} \textbf{\bibinfo{volume}{98}},
	\bibinfo{pages}{045136} (\bibinfo{year}{2018}).
	
	\bibitem{Wu2019d}
	\bibinfo{author}{Wu, Z.~M.} \emph{et~al.}
	\newblock \bibinfo{title}{{Multiple metamagnetism, extreme magnetoresistance
			and nontrivial topological electronic structures in the magnetic semimetal
			candidate holmium monobismuthide}}.
	\newblock \emph{\bibinfo{journal}{New J. Phys.}} \textbf{\bibinfo{volume}{21}},
	\bibinfo{pages}{093063} (\bibinfo{year}{2019}).
	
	\bibitem{Fente2013}
	\bibinfo{author}{Fente, A.} \emph{et~al.}
	\newblock \bibinfo{title}{{Low temperature magnetic transitions of single
			crystal HoBi}}.
	\newblock \emph{\bibinfo{journal}{Solid State Commun.}}
	\textbf{\bibinfo{volume}{171}}, \bibinfo{pages}{59--63}
	(\bibinfo{year}{2013}).
	
	\bibitem{Wada1995}
	\bibinfo{author}{Wada, H.}, \bibinfo{author}{Imai, H.} \&
	\bibinfo{author}{Shiga, M.}
	\newblock \bibinfo{title}{{Low temperature specific heat of DyBi and ErBi}}.
	\newblock \emph{\bibinfo{journal}{J. Alloys Compd.}}
	\textbf{\bibinfo{volume}{218}}, \bibinfo{pages}{73} (\bibinfo{year}{1995}).
	
	\bibitem{Ali2016}
	\bibinfo{author}{Ali, M.~N.} \emph{et~al.}
	\newblock \bibinfo{title}{{Butterfly magnetoresistance, quasi-2D Dirac Fermi
			surface and topological phase transition in ZrSiS}}.
	\newblock \emph{\bibinfo{journal}{Sci. Adv.}} \textbf{\bibinfo{volume}{2}},
	\bibinfo{pages}{e1601742} (\bibinfo{year}{2016}).
	
	\bibitem{Shekhar2015}
	\bibinfo{author}{Shekhar, C.} \emph{et~al.}
	\newblock \bibinfo{title}{{Extremely large magnetoresistance and ultrahigh
			mobility in the topological Weyl semimetal candidate NbP}}.
	\newblock \emph{\bibinfo{journal}{Nat. Phys.}} \textbf{\bibinfo{volume}{11}},
	\bibinfo{pages}{645} (\bibinfo{year}{2015}).
	
	\bibitem{Sankar2018}
	\bibinfo{author}{Sankar, R.} \emph{et~al.}
	\newblock \bibinfo{title}{{Crystal growth and transport properties of Weyl
			semimetal TaAs}}.
	\newblock \emph{\bibinfo{journal}{J. Phys. Condens. Matter}}
	\textbf{\bibinfo{volume}{30}}, \bibinfo{pages}{015803}
	(\bibinfo{year}{2018}).
	
	\bibitem{Singha2016a}
	\bibinfo{author}{Singha, R.}, \bibinfo{author}{Pariari, A.~K.},
	\bibinfo{author}{Satpati, B.} \& \bibinfo{author}{Mandal, P.}
	\newblock \bibinfo{title}{{Large nonsaturating magnetoresistance and signature
			of nondegenerate Dirac nodes in ZrSiS}}.
	\newblock \emph{\bibinfo{journal}{Proc. Natl. Acad. Sci. USA}}
	\textbf{\bibinfo{volume}{114}}, \bibinfo{pages}{2468} (\bibinfo{year}{2017}).
	
	\bibitem{Li2016f}
	\bibinfo{author}{Li, Y.} \emph{et~al.}
	\newblock \bibinfo{title}{{Resistivity plateau and negative magnetoresistance
			in the topological semimetal TaSb$_2$}}.
	\newblock \emph{\bibinfo{journal}{Phys. Rev. B}} \textbf{\bibinfo{volume}{94}},
	\bibinfo{pages}{121115} (\bibinfo{year}{2016}).
	
	\bibitem{Wang2015d}
	\bibinfo{author}{Wang, Y.~L.} \emph{et~al.}
	\newblock \bibinfo{title}{{Origin of the turn-on temperature behavior in
			WTe$_2$}}.
	\newblock \emph{\bibinfo{journal}{Phys. Rev. B}} \textbf{\bibinfo{volume}{92}},
	\bibinfo{pages}{180402} (\bibinfo{year}{2015}).
	
	\bibitem{Xu2017f}
	\bibinfo{author}{Xu, J.} \emph{et~al.}
	\newblock \bibinfo{title}{{Origin of the extremely large magnetoresistance in
			the semimetal YSb}}.
	\newblock \emph{\bibinfo{journal}{Phys. Rev. B}} \textbf{\bibinfo{volume}{96}},
	\bibinfo{pages}{075159} (\bibinfo{year}{2017}).
	
	\bibitem{Wu2015}
	\bibinfo{author}{Wu, Y.} \emph{et~al.}
	\newblock \bibinfo{title}{{Temperature-Induced Lifshitz Transition in
			WTe$_2$}}.
	\newblock \emph{\bibinfo{journal}{Phys. Rev. Lett.}}
	\textbf{\bibinfo{volume}{115}}, \bibinfo{pages}{166602}
	(\bibinfo{year}{2015}).
	
	\bibitem{Fan2020}
	\bibinfo{author}{Fan, L.-Y.} \emph{et~al.}
	\newblock \bibinfo{title}{{Anisotropic and extreme magnetoresistance in the
			magnetic semimetal candidate erbium monobismuthide}}.
	\newblock \emph{\bibinfo{journal}{Phys. Rev. B}}
	\textbf{\bibinfo{volume}{102}}, \bibinfo{pages}{104417}
	(\bibinfo{year}{2020}).
	
	\bibitem{Tafti2016a}
	\bibinfo{author}{Tafti, F.~F.} \emph{et~al.}
	\newblock \bibinfo{title}{{Temperature-field phase diagram of extreme
			magnetoresistance}}.
	\newblock \emph{\bibinfo{journal}{Proc. Natl. Acad. Sci. USA}}
	\textbf{\bibinfo{volume}{113}}, \bibinfo{pages}{E3475}
	(\bibinfo{year}{2016}).
	
	\bibitem{Wang2018}
	\bibinfo{author}{Wang, Y.} \emph{et~al.}
	\newblock \bibinfo{title}{{Topological semimetal state and field-induced Fermi
			surface reconstruction in the antiferromagnetic monopnictide NdSb}}.
	\newblock \emph{\bibinfo{journal}{Phys. Rev. B}} \textbf{\bibinfo{volume}{97}},
	\bibinfo{pages}{115133} (\bibinfo{year}{2018}).
	
	\bibitem{Hu2018b}
	\bibinfo{author}{Hu, Y.~J.} \emph{et~al.}
	\newblock \bibinfo{title}{{Extremely large magnetoresistance and the complete
			determination of the Fermi surface topology in the semimetal ScSb}}.
	\newblock \emph{\bibinfo{journal}{Phys. Rev. B}} \textbf{\bibinfo{volume}{98}},
	\bibinfo{pages}{035133} (\bibinfo{year}{2018}).
	
	\bibitem{Wu2018c}
	\bibinfo{author}{Wu, F.} \emph{et~al.}
	\newblock \bibinfo{title}{{Anomalous quantum oscillations and evidence for a
			non-trivial Berry phase in SmSb}}.
	\newblock \emph{\bibinfo{journal}{npj Quantum Mater.}}
	\textbf{\bibinfo{volume}{4}}, \bibinfo{pages}{20} (\bibinfo{year}{2019}).
	
	\bibitem{Zhao2022}
	\bibinfo{author}{Zhao, J.} \emph{et~al.}
	\newblock \bibinfo{title}{{Multiple metamagnetic transitions induced by high
			magnetic field in DyBi}}.
	\newblock \emph{\bibinfo{journal}{Phys. Rev. B}}
	\textbf{\bibinfo{volume}{106}}, \bibinfo{pages}{224412}
	(\bibinfo{year}{2022}).
	
	\bibitem{Wang2015}
	\bibinfo{author}{Wang, Y.~L.} \emph{et~al.}
	\newblock \bibinfo{title}{{Origin of the turn-on temperature behavior in
			${\mathrm{WTe}}_{2}$}}.
	\newblock \emph{\bibinfo{journal}{Phys. Rev. B}} \textbf{\bibinfo{volume}{92}},
	\bibinfo{pages}{180402} (\bibinfo{year}{2015}).
	
	\bibitem{Sun2016a}
	\bibinfo{author}{Sun, S.}, \bibinfo{author}{Wang, Q.}, \bibinfo{author}{Guo,
		P.-J.}, \bibinfo{author}{Liu, K.} \& \bibinfo{author}{Lei, H.}
	\newblock \bibinfo{title}{{Large magnetoresistance in LaBi: origin of
			field-induced resistivity upturn and plateau in compensated semimetals}}.
	\newblock \emph{\bibinfo{journal}{New J. Phys.}} \textbf{\bibinfo{volume}{18}},
	\bibinfo{pages}{082002} (\bibinfo{year}{2016}).
	
	\bibitem{Qian2018}
	\bibinfo{author}{Qian, B.} \emph{et~al.}
	\newblock \bibinfo{title}{{Extremely large magnetoresistance in the nonmagnetic
			semimetal YBi}}.
	\newblock \emph{\bibinfo{journal}{J. Mater. Chem. C}}
	\textbf{\bibinfo{volume}{6}}, \bibinfo{pages}{10020--10029}
	(\bibinfo{year}{2018}).
	
	\bibitem{Wang2017b}
	\bibinfo{author}{Wang, A.} \emph{et~al.}
	\newblock \bibinfo{title}{{Large magnetoresistance in the type-II Weyl
			semimetal WP$_2$}}.
	\newblock \emph{\bibinfo{journal}{Phys. Rev. B}} \textbf{\bibinfo{volume}{96}},
	\bibinfo{pages}{121107} (\bibinfo{year}{2017}).
	
	\bibitem{Yang2017c}
	\bibinfo{author}{Yang, H.-Y.} \emph{et~al.}
	\newblock \bibinfo{title}{{Extreme magnetoresistance in the topologically
			trivial lanthanum monopnictide LaAs}}.
	\newblock \emph{\bibinfo{journal}{Phys. Rev. B}} \textbf{\bibinfo{volume}{96}},
	\bibinfo{pages}{235128} (\bibinfo{year}{2017}).
	
	\bibitem{Ziman2001}
	\bibinfo{author}{Ziman, J.}
	\newblock \emph{\bibinfo{title}{Electrons and Phonons: The Theory of Transport
			Phenomena in Solids}} (\bibinfo{publisher}{Oxford University Press},
	\bibinfo{year}{2001}).
	
	\bibitem{Pavlosiuk2020}
	\bibinfo{author}{Pavlosiuk, O.}, \bibinfo{author}{Fałat, P.},
	\bibinfo{author}{Kaczorowski, D.} \& \bibinfo{author}{Wiśniewski, P.}
	\newblock \bibinfo{title}{{Anomalous Hall effect and negative longitudinal
			magnetoresistance in half-Heusler topological semimetal candidates TbPtBi and
			HoPtBi}}.
	\newblock \emph{\bibinfo{journal}{APL Mater.}} \textbf{\bibinfo{volume}{8}},
	\bibinfo{pages}{111107} (\bibinfo{year}{2020}).
	
	\bibitem{Xu2021d}
	\bibinfo{author}{Xu, J.} \emph{et~al.}
	\newblock \bibinfo{title}{{Extended Kohler's Rule of Magnetoresistance}}.
	\newblock \emph{\bibinfo{journal}{Phys. Rev. X}} \textbf{\bibinfo{volume}{11}},
	\bibinfo{pages}{041029} (\bibinfo{year}{2021}).
	
	\bibitem{Xie2020}
	\bibinfo{author}{Xie, W.} \emph{et~al.}
	\newblock \bibinfo{title}{{Magnetotransport and electronic structure of the
			antiferromagnetic semimetal YbAs}}.
	\newblock \emph{\bibinfo{journal}{Phys. Rev. B}}
	\textbf{\bibinfo{volume}{101}}, \bibinfo{pages}{085132}
	(\bibinfo{year}{2020}).
	
	\bibitem{Kakihana2019}
	\bibinfo{author}{Kakihana, M.} \emph{et~al.}
	\newblock \bibinfo{title}{{Fermi Surface Properties of Semimetals YSb, LuSb,
			YBi, and LuBi Studied by the de Haas–van Alphen Effect}}.
	\newblock \emph{\bibinfo{journal}{J. Phys. Soc. Jpn.}}
	\textbf{\bibinfo{volume}{88}}, \bibinfo{pages}{044712}
	(\bibinfo{year}{2019}).
	
	\bibitem{Hosen2020a}
	\bibinfo{author}{Hosen, M.~M.} \emph{et~al.}
	\newblock \bibinfo{title}{{Observation of gapped state in rare-earth
			monopnictide HoSb}}.
	\newblock \emph{\bibinfo{journal}{Sci. Rep.}} \textbf{\bibinfo{volume}{10}},
	\bibinfo{pages}{12961} (\bibinfo{year}{2020}).
	
	\bibitem{Sakhya2021a}
	\bibinfo{author}{Sakhya, A.~P.}, \bibinfo{author}{Paulose, P.~L.},
	\bibinfo{author}{Thamizhavel, A.} \& \bibinfo{author}{Maiti, K.}
	\newblock \bibinfo{title}{{Evidence of nontrivial Berry phase and Kondo physics
			in SmBi}}.
	\newblock \emph{\bibinfo{journal}{Phys. Rev. Materials}}
	\textbf{\bibinfo{volume}{5}}, \bibinfo{pages}{054201} (\bibinfo{year}{2021}).
	
	\bibitem{Wang2018b}
	\bibinfo{author}{Wang, Y.-Y.}, \bibinfo{author}{Sun, L.-L.},
	\bibinfo{author}{Xu, S.}, \bibinfo{author}{Su, Y.} \& \bibinfo{author}{Xia,
		T.-L.}
	\newblock \bibinfo{title}{{Unusual magnetotransport in holmium
			monoantimonide}}.
	\newblock \emph{\bibinfo{journal}{Phys. Rev. B}} \textbf{\bibinfo{volume}{98}},
	\bibinfo{pages}{045137} (\bibinfo{year}{2018}).
	
	\bibitem{Wang2018a}
	\bibinfo{author}{Wang, Y.} \emph{et~al.}
	\newblock \bibinfo{title}{{Topological semimetal state and field-induced Fermi
			surface reconstruction in the antiferromagnetic monopnictide NdSb}}.
	\newblock \emph{\bibinfo{journal}{Phys. Rev. B}} \textbf{\bibinfo{volume}{97}},
	\bibinfo{pages}{115133} (\bibinfo{year}{2018}).
	
	\bibitem{Pavlosiuk2018c}
	\bibinfo{author}{Pavlosiuk, O.}, \bibinfo{author}{Kleinert, M.},
	\bibinfo{author}{Wi{\'{s}}niewski, P.} \& \bibinfo{author}{Kaczorowski, D.}
	\newblock \bibinfo{title}{{Antiferromagnetic Order in the Half-Heusler Phase
			TbPdBi}}.
	\newblock \emph{\bibinfo{journal}{Acta Phys. Pol. A}}
	\textbf{\bibinfo{volume}{133}}, \bibinfo{pages}{498} (\bibinfo{year}{2018}).
	
	\bibitem{Budko2008}
	\bibinfo{author}{Bud'ko, S.~L.} \emph{et~al.}
	\newblock \bibinfo{title}{{Thermal expansion and magnetostriction of pure and
			doped RAgSb$_2$ (R = Y, Sm, La) single crystals}}.
	\newblock \emph{\bibinfo{journal}{J. Phys. Condens. Matter}}
	\textbf{\bibinfo{volume}{20}}, \bibinfo{pages}{115210}
	(\bibinfo{year}{2008}).
	
	\bibitem{Guo2021a}
	\bibinfo{author}{Guo, C.} \emph{et~al.}
	\newblock \bibinfo{title}{{Temperature dependence of quantum oscillations from
			non-parabolic dispersions}}.
	\newblock \emph{\bibinfo{journal}{Nat. Commun.}} \textbf{\bibinfo{volume}{12}},
	\bibinfo{pages}{6213} (\bibinfo{year}{2021}).
	
	\bibitem{Jiang2021b}
	\bibinfo{author}{Jiang, Q.} \emph{et~al.}
	\newblock \bibinfo{title}{{Quantum oscillations in the field-induced
			ferromagnetic state of MnBi$_{2\mbox{-}x}$Sb$_x$Te$_4$}}.
	\newblock \emph{\bibinfo{journal}{Phys. Rev. B}}
	\textbf{\bibinfo{volume}{103}}, \bibinfo{pages}{205111}
	(\bibinfo{year}{2021}).
	
	\bibitem{Lyu2020a}
	\bibinfo{author}{Lyu, M.} \emph{et~al.}
	\newblock \bibinfo{title}{{Nonsaturating magnetoresistance, anomalous Hall
			effect, and magnetic quantum oscillations in the ferromagnetic semimetal
			PrAlSi}}.
	\newblock \emph{\bibinfo{journal}{Phys. Rev. B}}
	\textbf{\bibinfo{volume}{102}}, \bibinfo{pages}{085143}
	(\bibinfo{year}{2020}).
	
	\bibitem{Goll2006}
	\bibinfo{author}{Goll, G.} \emph{et~al.}
	\newblock \bibinfo{title}{{Temperature-dependent Fermi surface in CeBiPt}}.
	\newblock \emph{\bibinfo{journal}{Europhys. Lett.}}
	\textbf{\bibinfo{volume}{57}}, \bibinfo{pages}{233} (\bibinfo{year}{2002}).
	
	\bibitem{Rodriguez-Carvajal1993}
	\bibinfo{author}{Rodr{\'{i}}guez-Carvajal, J.}
	\newblock \bibinfo{title}{{Recent advances in magnetic structure determination
			by neutron powder diffraction}}.
	\newblock \emph{\bibinfo{journal}{Phys. B: Condens. Matter}}
	\textbf{\bibinfo{volume}{192}}, \bibinfo{pages}{55--69}
	(\bibinfo{year}{1993}).
	
	\bibitem{Yoshihara1975}
	\bibinfo{author}{Yoshihara, K.}, \bibinfo{author}{Taylor, J.~B.},
	\bibinfo{author}{Calvert, L.~D.} \& \bibinfo{author}{Despault, J.~G.}
	\newblock \bibinfo{title}{Rare-earth bismuthides}.
	\newblock \emph{\bibinfo{journal}{{J. Less-Common Met.}}}
	\textbf{\bibinfo{volume}{41}}, \bibinfo{pages}{329--337}
	(\bibinfo{year}{1975}).
	
\end{thebibliography}

\end{document}